\documentclass[12pt,preprint]{aastex}

\slugcomment{}

\begin{document}

\title{High-Latitude Molecular Clouds as $\gamma$-ray Sources for GLAST}

\author{Diego F. Torres$^1$, T. M. Dame$^2$,  \& Seth W. Digel$^3$
\\ {\small $^1$ Lawrence Livermore National Laboratory,  7000 East
Ave., L-413, Livermore CA 94550, USA. E-mail:
dtorres@igpp.ucllnl.org}\\
{\small $^2$ Harvard-Smithsonian Center for Astrophysics, 60 Garden St., MS72,
Cambridge, MA 02138, USA.  E-mail: tdame@cfa.harvard.edu} \\
{\small $^3$ Stanford Linear Accelerator Center, 2575 Sand Hill
Road, M/S 43A, Menlo Park, CA  94025, USA.  E-mail: digel@slac.stanford.edu}}

\begin{abstract}

For about two decades, a population of relative small and nearby
molecular clouds has been known to exist at high Galactic latitudes.
Lying more than 10$^\circ$ from the Galactic plane, these clouds
have typical distances of $\sim$150 pc, angular sizes of
$\sim$1$^\circ$, and masses of order tens of solar masses. These
objects are passive sources of high-energy $\gamma$-rays through
cosmic ray-gas interactions. Using a new wide-angle CO survey of the
northern sky, we show that typical high-latitude clouds are not
bright enough in $\gamma$-rays to have been detected by EGRET, but
that of order 100 of them will be detectable by the Large Area
Telescope (LAT) on GLAST.  Thus, we predict a new steady population
of $\gamma$-ray sources at high Galactic latitudes, perhaps the most
numerous after active galactic nuclei.

\end{abstract}

\section{Introduction}

The great majority of $\gamma$-ray point sources at high Galactic
latitudes in the Third Energetic $\gamma$-ray Experiment Telescope
(EGRET) catalog is active galactic nuclei (AGNs, Hartman et al.
1999). This is also the expectation for the sources that will be
detected by the  $\gamma$-ray Large Area Space Telescope (GLAST)
mission, to be launched by NASA in early 2007. However, Punsly
(1997) and Sowards-Emmerd et al. (2002; 2004), among others, have
already noted that ascribing all EGRET detections at high latitudes
to AGNs is not possible. Numerous other potential $\gamma$-ray
sources at high latitudes have been suggested, including nearby
starburst galaxies (e.g., Paglione et al. 1996), luminous infrared
galaxies (Torres et al. 2004a; Torres 2004b), normal galaxies (e.g.,
Pavlidou \& Fields 2001), clouds of baryonic dark matter in the
Galactic halo (Walker et al. 2003), galaxy clusters (e.g., Reimer et
al. 2003 and references therein), and radio galaxies (e.g. Mukherjee
et al. 2002; Combi et al. 2003), but not all of them are viable (see
the review by Torres 2004c).

In this Letter we show that another significant population of
high-latitude $\gamma$-ray sources will be relevant for the next
generation $\gamma$-ray telescopes: relatively small and nearby
molecular clouds. The $\gamma$-ray fluxes from these objects depend
only on their angular sizes and molecular column densities---quantities
that can be inferred by CO surveys---and on the ambient
cosmic ray (CR) spectrum. Since molecular clouds lie in such a thin
layer ($\sim$87 pc HWHM; Dame et al. 1987), nearly all clouds at
$|b| > 10^\circ$ will lie within 1 kpc of the Sun, and many will be
much closer than that. We can therefore assume that the ambient CR
spectrum is the same as that measured near the Earth. Using the
instrumental characteristics of the Large Area Telescope (LAT) on
GLAST, we estimate here that this instrument will be able to detect of
order 100 high-latitude clouds.  Since these clouds will be steady
$\gamma$-ray sources, variability indices (Torres et al. 2001; Nolan
et al. 2003) will be useful to distinguish them from AGNs.

The $\gamma$-ray emission from high-latitude clouds is of interest
not only for what it can reveal about the masses of these objects
and the local CR spectrum, but also as a potential source of
confusion for identifying other types of $\gamma$-ray sources
detected by the LAT. As was the case for EGRET, the study of the
extragalactic sky with the LAT will require an accurate model for
the diffuse $\gamma$-ray emission produced by CR interactions with
the Galactic interstellar medium. Small molecular clouds not
included in the diffuse model, or simply assigned incorrect column
densities, might be interpreted as point sources. The lack of CO
surveys over large areas of the high-latitude sky was not a serious
problem for modeling the diffuse emission detected by EGRET (Bertsch
et al. 1993, Hunter et al. 1997) because it was evident even from
optical and far-infrared surveys that no molecular clouds large
enough to be detected with the limited sensitivity and resolution of
EGRET existed in these areas. Interpretation of the much more
sensitive and higher angular resolution LAT survey will require a
far more accurate accounting of the gamma rays from molecular gas at
high latitudes.

\section{Surveys of high-latitude clouds}

The molecular gas at $|b| > 10^\circ$ consists primarily of
translucent clouds\footnote{Small molecular clouds are divided
into three classes on the basis of their visual extinction and
astrochemical properties (van Dishoeck \& Black 1988): diffuse,
translucent, and dark. Translucent clouds, the intermediate regime,
have CO abundances in the range $10^{-6}-10^{-4}$ and column
densities greater than $10^{15}$ cm$^{-2}$. Different
from dark clouds, where the chemistry is driven by collisional
processes, translucent clouds are still dominated by photoprocesses.
It is in this translucent regime where most of the carbon becomes
molecular, and so they are detectable in CO, but are relatively
optically thin -- and thus their name -- exhibiting low optical
extinctions. }, although a few more massive objects are found in this
region, most notably the giant molecular cloud in Orion (Wilson et
al. 2004), the Polaris Flare\footnote{A large molecular cloud with
a mass of 5500 M$_\odot$ at a distance of $\sim 100$ pc (Meyerdierks
\& Heithausen 1996 and references therein); its $\gamma$-ray
emission was analyzed by Digel et al. (1996).} and the dark clouds
in Taurus (Ungerechts \& Thaddeus 1987) and Ophiuchus (de Geus et
al. 1990).

The first large-scale search for high-latitude molecular clouds was
carried out by Magnani, Blitz \& Mundy (MBM, 1985), who surveyed CO
toward regions of optical obscuration on the Palomar Observatory Sky
Survey plates.  MBM reported finding 57 clouds in 35 complexes at
$|b|>25^\circ$, distributed asymmetrically with respect to the
Galactic plane, at typical distances of 100 pc. More clouds were
found in the southern Galactic hemisphere than in the north, a fact
interpreted as a displacement of the position of the Sun from the
midplane of the Galaxy. This work was updated by Magnani, Hartmann
\& Speck (MHS, 1996), who compiled a catalog of molecular clouds at
high Galactic latitudes consisting of 120 members.

Table 3 of MHS details the average properties of the clouds in their
catalog. The distances range from 60 to 350 pc, with a typical value
around 150 pc, whereas the masses range from $\cal O$(0.1) to $\cal
O$(100) M$_\odot$. The mean values for mass and H$_2$ density are 40
M$_\odot$ (46 M$_\odot$ if the Polaris flare is included) and 140
cm$^{-3}$, respectively. The scale height inferred from the velocity
dispersion of the clouds was 124 pc and the average solid angle per
cloud was ~0.8 square degrees. The filling fraction of the sample,
averaged over the whole sky at $|b|>25^\circ$, is $\sim 0.005$.

More complete surveys in CO have been conducted both for the
northern (Hartmann et al. 1998) and southern (Magnani et al. 2000)
Galactic hemispheres at $|b|>30^\circ$, although the gridding of the
searches was still coarse (about 1$^\circ$) and allowed for some
clouds to be missed (see below). In the northern hemisphere, only
2 molecular clouds previously unknown were found. In contrast,
58 new detections were reported in the southern hemisphere survey,
and another 75 were found to be related with 26 previously
catalogued clouds situated within the survey boundaries. Using the
southern survey, the filling factor of molecular gas was found to be
0.03, an order of magnitude larger than that reported using the
survey of the northern hemisphere. Average results concerning
masses, densities, and distances were not significantly different
from those of MBM.

Over the past 3 years, the Galactic CO ($J$ = 1--0) survey of Dame,
Hartmann, \& Thaddeus (2001; hereafter DHT) has been extended with
the CfA 1.2 meter telescope to cover all of the area at $|b| <
30^\circ$ and ${\delta}> -17^\circ$ ($l  <  230^\circ$) with a
sampling interval of 1/4$^\circ$ or better (Dame \& Thaddeus 2004).
These new observations have found more than 200 relatively small and
isolated molecular clouds more than $10^\circ$ from the plane.
Since this new survey covers a large fraction of the Galaxy with a
resolution and sensitivity to molecular clouds that exceeds that of
the LAT, we will use it in Section 4 to estimate the total number of
high-latitude clouds that the LAT will detect.

\section{$\gamma$-ray emission from clouds and detectability}

With a cloud considered as a passive target for CR interactions,
the hadronically-generated $\gamma$-ray number luminosity (photons
per unit time) can be computed as (see, e.g., Aharonian 2001, Torres
et al. 2003) $ L_{\gamma}(E_{\gamma})=\int n(r) q_{\gamma}(E_\gamma)
dV \sim ({M}/{m_p}) {q_\gamma}, $ where $ r$ represents the position
within the interaction region $V$, $M$ is the mass of gas, $m_p$ is
the proton mass, $n$ is the number density, and ${q_\gamma}$ is the
$\gamma$-ray emissivity (photons per unit of time per atom). The
$\gamma$-ray flux is then $F(>100\, {\rm MeV})=L_\gamma(>100 \, {\rm
MeV})/4\pi {D}^2$, where $D$ is the distance to the cloud. In an
appropriate scaling, the flux is $ F(>100\; {\rm MeV}) \sim 2.4
\times 10^{-9} ( {M}/{10 {\rm M}_\odot}) ( {D}/{100 {\rm pc}} )^{-2}
k \;{\rm photons\; cm^{-2}~s^{-1}}, $  where $k$ is the enhancement
factor of CRs.  The numerical factor takes into account electron
bremsstrahlung (see, e.g., Pavlidou \& Fields 2001). For local high-latitude clouds,
we assume no enhancement, $k=1$, so that the CR spectrum is the same
as the proton flux measured in the neighborhood of the Earth.


The molecular masses of each cloud can be estimated under the
usual assumption of a
proportionality between the velocity-integrated CO intensity $W_{\rm
CO}$ and the H$_2$ column density,
$N({\rm{H_2}})$, $X\equiv N({\rm{H_2}})/W_{\rm CO}$.
We use the value $X = 1.8 \times 10^{20}$ cm$^{-2}$ (K km
s$^{-1})^{-1}$, derived from an intercomparison of large-scale
far-infrared, 21 cm, and CO surveys (DHT).\footnote{Recently,
Magnani et al. (2003), although favoring the use of the $X$-factor
for population analyses like ours, provided evidence that in
translucent clouds, CO is not a linear tracer of the column density
of molecular hydrogen. They showed that for at least one cloud, the
CO/H$_2$ ratio varies as a function of the molecular hydrogen column
density.  Magnani et al. (2003) proposed to use CH observations to
calibrate the CO-to-H$_2$ conversion factor. Given the fact that
these observations are not available for all clouds (not even for a
significant fraction of them) we continue using the standard
approach.} The value of $X$ is uncertain for translucent clouds:
other works have adopted values that are either larger or smaller,
by at least a factor of 2 than the value adopted here
(e.g., de Vries, Heithausen \& Thaddeus 1987; Magnani \& Onello 1995,
Hartmann et al. 1998). Thus, the mass estimate should perhaps be
considered uncertain by a comparable factor.

Expressed in terms of its integrated intensity in the CO line,
$S_{\rm CO}$, the mass of molecular hydrogen is $M [{\rm M}_\odot]=
860 S_{\rm CO}[{\rm K\; km\, s^{-1}\, deg^2}] D[{\rm kpc}]^2$ (Dame
et al. 1986). The factor 860 incorporates the value of $X$
assumed in this paper; the correction for the contributions from
helium and heavier elements is removed here, because the
$\gamma$-ray emissivity that we use (see below) already corrects for
their contributions to the overall cross sections by relative
abundance.  Note that since the $\gamma$-ray flux from a cloud
scales as $M/{D}^2$ and $M$ scales as $S_{\rm CO}D^2$, the
$\gamma$-ray flux is linearly proportional to $S_{\rm CO}$, which is
a directly observable quantity.

In estimating cloud masses, we do not include atomic hydrogen
associated with the molecular gas.  The atomic gas is typically more
smoothly distributed than the molecular gas on angular scales of
degrees. Effectively we are considering the diffuse $\gamma$-ray
emission from the pervasive atomic hydrogen as part of the Galactic
background against which the molecular clouds are detected.

In order to determine which clouds the LAT will be able to detect
and spatially resolve, we simulated observations of idealized,
disk-shaped sources against an isotropic background that includes
contributions from the extragalactic sky as well as cosmic-ray
interactions in Galactic H~I and inverse Compton scattering of the
interstellar radiation field.  The background intensity was $2
\times 10^{-5}$ cm$^{-2}$  s$^{-1}$ sr$^{-1}$, $>100$ MeV, photon
spectral index $-2.1$), typical of the Milky Way at high latitude.
We simulated a 1-year sky survey of the sort planned for the first
year after instrument checkout. These simulations are shown in
Figure 1. The sources were assumed to have a spectrum consistent
with the local $\gamma$-ray emissivity (primarily from $\pi^0$ decay
in the range $>$100 MeV, Bertsch et al. 1993). Preliminary response
functions for the LAT, consistent with the performance requirements
(GLAST Science Requirements Doc. 2003), were used.  The actual
limits and angular sizes can be expected to change somewhat when
final response functions become available.

As Figure 1 illustrates, the criterion for detectability by the LAT
cannot be based solely  on expected flux.  For faint extended
$\gamma$-ray sources, detectability by the LAT will depend rather
strongly on angular size (see below).

In the following we use `resolve' to mean determine to 5$\sigma$ or
better that the source is inconsistent with being point-like.
Resolving a source requires better statistics (i.e., more
$\gamma$-rays) than detecting it; thus the flux limit for detecting
a source is lower than the limit for resolving it.  Again using
simulations with disk-shaped clouds that have $\gamma$-ray spectra
consistent with the local emissivity, we mapped the regions of
detectability and resolvability in the flux-angular size plane. The
result is shown in Figure 2$a$, where we have also overlaid all of the
molecular clouds in the range $|b| = 10^\circ - 30^\circ, l =
0^\circ - 230^\circ$ detected by the recently-extended DHT survey
(see Section 4).

Sources in the lightest shaded region of Figure 2$a$ are both
detectable and resolvable, i.e., distinguishable from point sources.
In the intermediate shaded region sources are detectable but not
distinguishable from point sources.  Even if the expected location,
extent, and shape of a source (disks in the case of the simulations)
are known, if it lies within this region it is indistinguishable
from a point source.  Figure 2$a$ indicates that at high latitudes a
molecular cloud with flux as great as twice the detection limit for
small clouds will not be detectable if its angular diameter,
$\beta$, is greater than $\sim$1$^\circ$. Conversely, many of the
clouds that are detectable will not be resolvable (i.e., not
distinguishable from point sources). In particular any detectable
cloud with flux less than $\sim$8 $\times 10^{-9}$ cm$^{-2}$
s$^{-1}$ ($>$100 MeV) will not be resolvable.

Resolving a cloud of course unambiguously establishes it as a diffuse source.
The diffuse nature of the clouds might also be inferred from
variability studies -- they should be steady -- or from their relatively soft spectra at high
energies (the local $\gamma$-ray emissivity falling as $\sim
E^{-2.75}$ in the GeV range).  However, variability and spectral
studies for faint sources such as these small clouds will not be particularly constraining.
We expect that many of these $\gamma$-ray sources would be unidentified
if they were not already cataloged as molecular clouds.

\begin{figure}[t]
 \begin{center}
 \includegraphics[width=.45\textwidth]{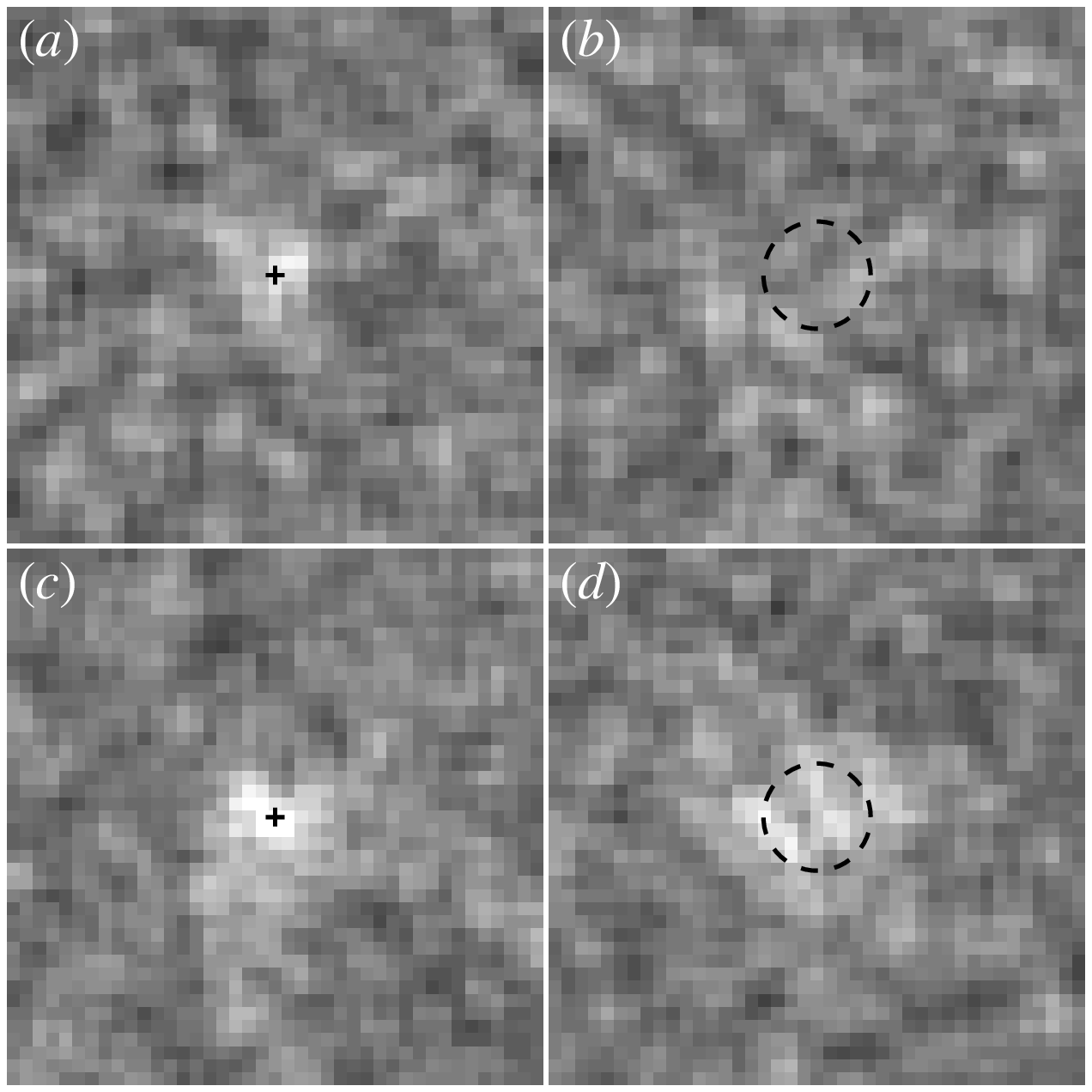}
\caption{Illustration of the limits of the resolving
power of the LAT.  The images are simulated $\gamma$-ray intensity
maps (4$^\circ \times$ 4$^\circ$) for LAT observations of isolated
sources against the diffuse background intensity at high latitudes.
The simulation is based on preliminary response functions for the
LAT, the exposure is that of the planned 1-year scanning survey of
the sky, and the sources are assumed to have spectra consistent with
the local $\gamma$-ray emissivity (see text). ($a$) Point source
with flux $5 \times 10^{-9}$ cm$^{-2}$ s$^{-1}$ ($>$100 MeV). The
position of the source is indicated by the cross. ($b$) Disk-shaped
source with 1$^\circ$ angular diameter and the same flux as ($a$).
The circle indicates the position and extent of the source.  The
source is not detectable. ($c$-$d$) are as for ($a$-$b$) except with
a source flux of $1 \times 10^{-8}$ cm$^{-2}$ s$^{-1}$ ($>$100 MeV).
The images have been smoothed slightly to reduce statistical
fluctuations; each 0.1$^\circ \times$ 0.1$^\circ$ pixel typically
contains
 2 $\gamma$-rays.}  \label{lat}
 \end{center}
\end{figure}

\section{The total number of LAT-detectable clouds}

In this section we use the CfA Galactic CO survey, including the
recent large extension by Dame \& Thaddeus (2004), to estimate the
total number of clouds at  $|b| > 10^\circ$ that will be detected by
the LAT, and the number of these that will be resolved. We generated
a velocity-integrated CO map from the survey using the masked-moment
method described in DHT. Defining individual clouds as isolated
regions of statistically significant emission, we derived a catalog
of 232 clouds in the region ($|b| = 10^\circ - 30^\circ$, $l <
230^\circ$) with $S_{\rm CO}$ ranging from 0.09 to 1418 ${\rm K\;
km\, s^{-1}\, deg^2}$. The minimum value corresponds to a
$\gamma$-ray flux of $0.2 \times 10^{-9} {\rm photons\;
cm^{-2}~s^{-1}}$ (see Sec. 3), which as Figure 2$a$ shows is well
below the detection threshold for the LAT; the corresponding mass is
$1.7$ ${\rm M}_\odot$ at 150 pc. The fluxes and angular diameters of
the clouds are indicated in Figure 2$a$ as crosses.  Many of the
cataloged clouds lie below the flux range of the figure and 20 lie
above it (see below).


This catalog certainly contains all clouds in the survey area that
might be detectable by the LAT, and it covers such a large range of
Galactic longitude and latitude that the total number of clouds---or
the number in any given range of $S_{\rm CO}$ or size---can be
reliably extrapolated to the whole sky at $|b| > 10^\circ$.  The
number of cataloged clouds per unit solid angle closely follows a
cosecant dependence on latitude, as expected for clouds in a
plane-parallel layer. On the assumption that this dependence extends
to $90^\circ$ latitude and that there is no systematic dependence on
longitude, the extrapolation factor from the survey area to the
whole sky at $|b| > 10^\circ$ is 2.8.

As noted above, under the assumption of a uniform local CR flux and
the $X$-ratio of DHT, the detectability of a local molecular cloud by the
LAT depends on its total CO flux $S_{\rm CO}$ and its angular
size $\beta$.  Figure 2$b$ shows the total number of
clouds as a function of minimum $S_{\rm CO}$, extrapolated from the
survey area to the whole sky at $|b| > 10^\circ$.
Figure 2$c$ shows the same for cloud angular diameters.

A total of 53 clouds in the survey area are expected to be
detectable by the LAT; this extrapolates to 143 such clouds over the
whole sky at $|b| > 10^\circ$. (Figure 2$a$ does not show the 20
with the greatest fluxes and diameters, which are sparsely
distributed outside the range of the plot.) A smaller but still
significant number of clouds will not only be detected but also
resolved: 30 over the survey area or 78 over the whole sky at $|b| >
10^\circ$. Since many clouds at high latitude cluster into larger associations that follow HI filaments or shells (Gir, Blitz, \& Magnani 1994; see also Yamamoto et al. 2003), some otherwise undetectable small clouds likely will form larger structures that will be resolved by the LAT.

\begin{figure}[t]
\begin{center}
%
\includegraphics[width=.32\textwidth,height=4cm]{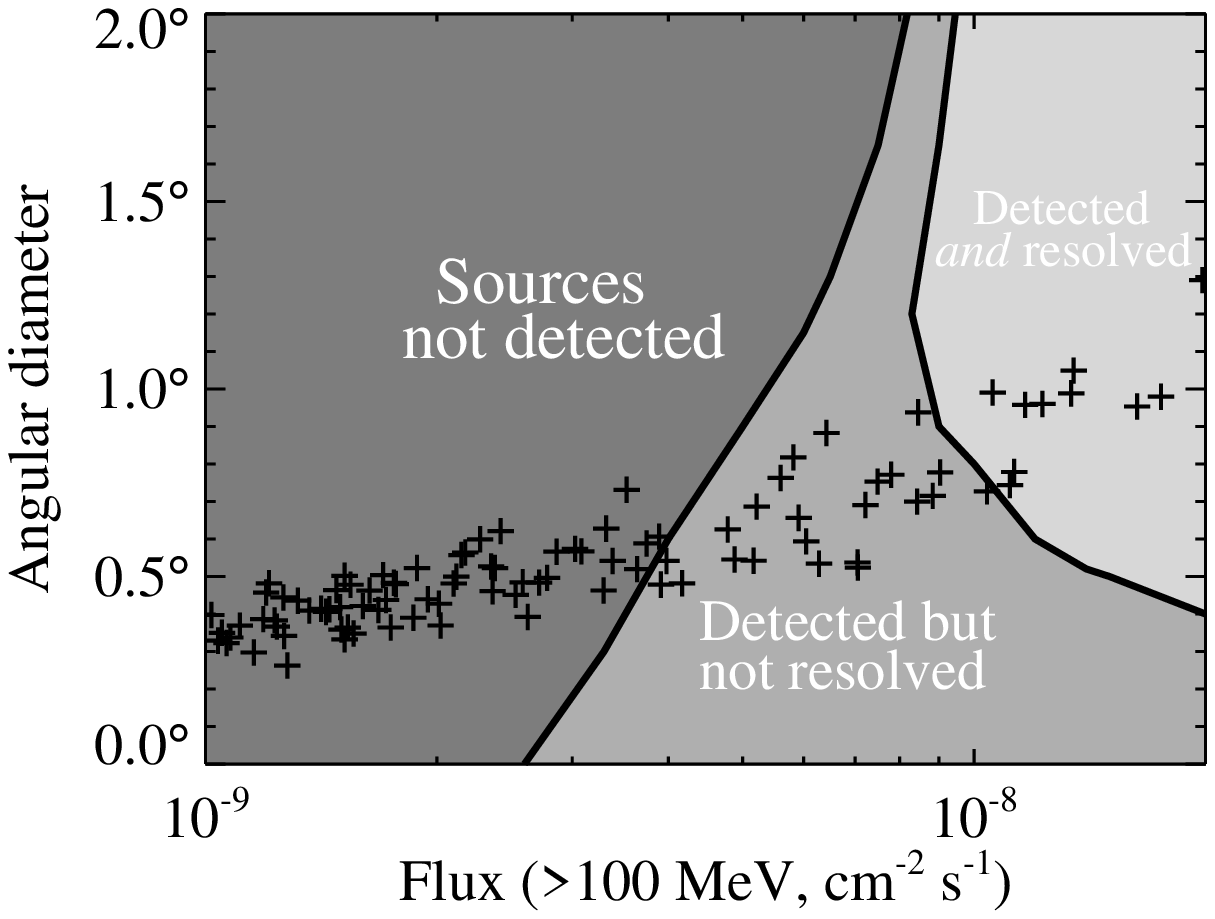}
\includegraphics[width=.32\textwidth,height=4cm]{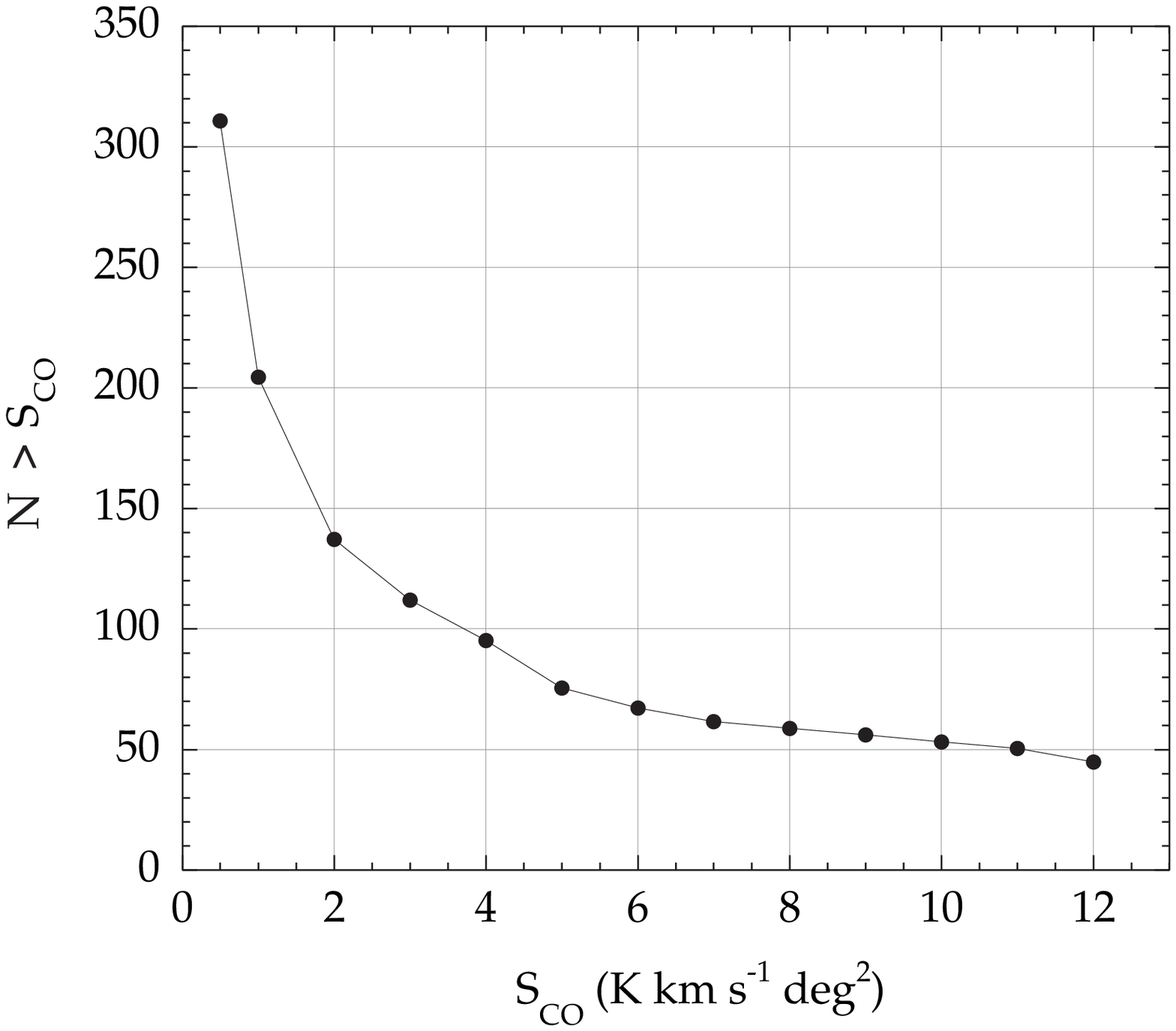}
\includegraphics[width=.32\textwidth,height=4cm]{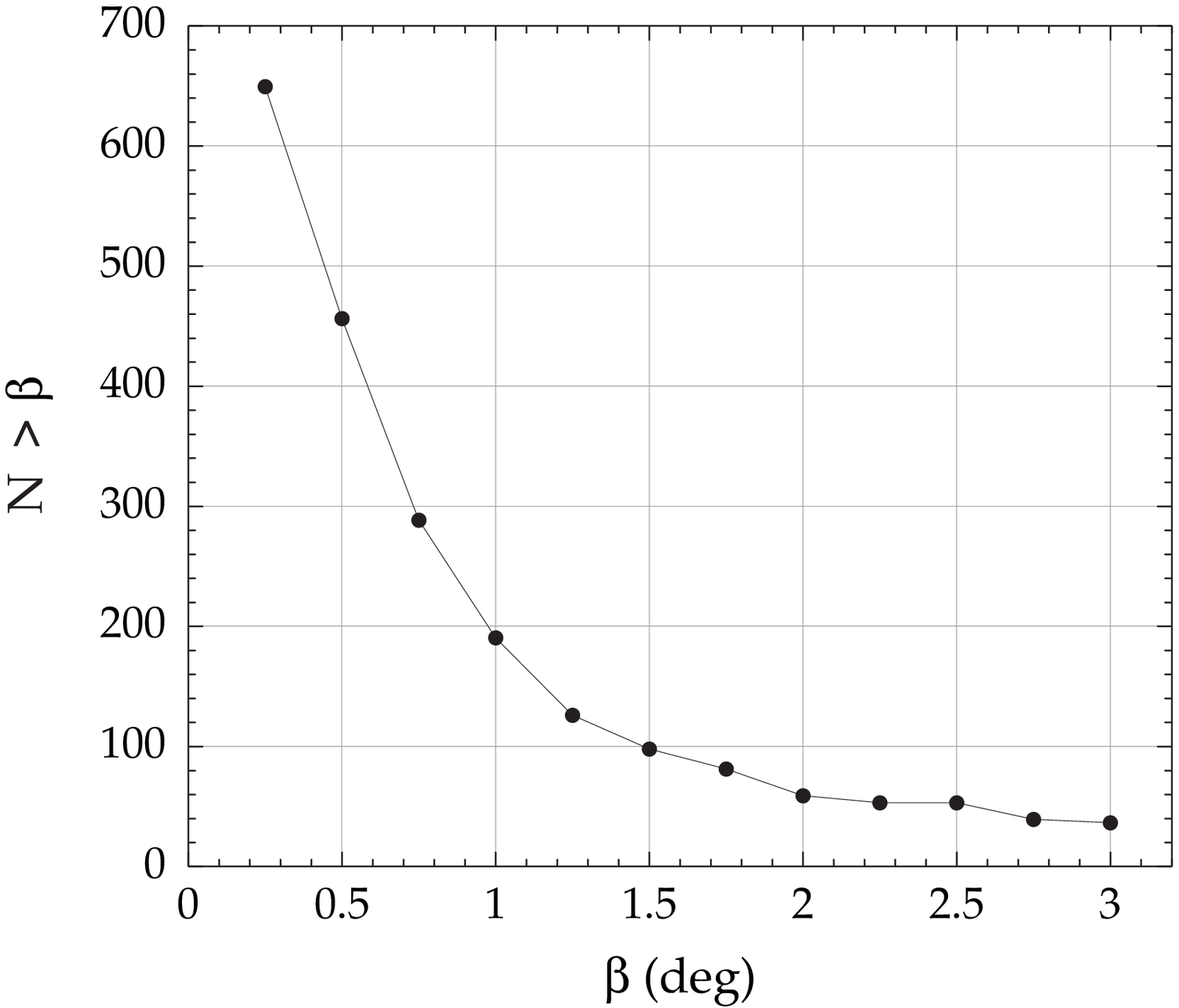}
\caption{($a$) The limiting fluxes for sources with a given angular
diameter to be detected and resolved by LAT.  See text for details.
The heavy-drawn lines indicate the boundaries for detecting and/or
resolving a source at the 5$\sigma$ level.  The boundaries were
derived from simulations using preliminary response functions for
the LAT, for a 1-yr sky survey.  The catalog contains 20 clouds that
have greater fluxes and angular diameters than the limits of this
plot (and many others below the flux range of the Figure).  These
largest clouds all will be detected and resolved by the LAT. The
other panels show the extrapolated total numbers of clouds at $|b| >
10^\circ$ as a function of minimum $S_{\rm CO}$ (b) and minimum
angular diameter (c); see Section 4. } \label{TOM1}
\end{center}
\end{figure}


\section{Concluding remarks}


It is important to understand and to account for the $\gamma$-ray flux from
high-latitude molecular clouds both as individual sources and as
contributors to the diffuse background. Unlike the proposed associations
between low-latitude EGRET sources and clouds adjacent to
supernova remnants (e.g., Romero et al. 1999), local high-latitude
clouds are passive targets for CR interactions. Except for those
those lying very close to stellar OB associations (Bhat 2000), we
expect them to be subject to the same CR spectrum as
that found locally.

Detecting these clouds will provide information on the local CR
spectrum, particularly in cases where the cloud distance is well
determined (e.g., Luhman 2001, Grant \& Burrows 1999). In addition,
if as was done with EGRET observations at low latitudes, H I data
are used to calibrate the $\gamma$-ray emissivity, comparison of
cloud  $\gamma$-ray fluxes with CO luminosities will provide not
only a calibration of the $X$-factor, but also information on
possible variations of $X$ with cloud properties or Galactic
location.

The present results underscore the need for continued large-scale
mapping of CO at medium and high Galactic latitudes. Only in the
mid-latitude region analyzed here ($|b| = 10-30^\circ$, $l < 230^\circ$;
Dame \& Thaddeus 2004) can we claim that most LAT-detectable clouds
have been found, and even in this region more finely-sampled
CO maps are required to accurately determine cloud sizes and masses.
In the southern Galaxy CO observations more than a few degrees from
the plane are still extremely sparse.

While the analysis regarding detectability/resolvability presented
here (see Figure 1) is applicable to all possible $\gamma$-ray
sources, our investigation shows that nearby molecular clouds may in
fact be the most numerous population of sources other than that of
blazars that the LAT will detect at high latitudes.

\section*{Acknowledgments}

The work of DFT was performed under the auspices of the U.S. DOE by
UC's LLNL under contract No. W-7405-Eng-48, and it was in part done
while visiting the Institut de F\'{\i}sica d'Altes Energies, Spain.
He acknowledges its hospitality.

\end{document}